\documentclass[prl,twocolumn]{revtex4}

\usepackage{graphicx}

\begin{document}


\title{Two-Level Laser-like Emission by the Interaction of Self-Induced Transparency Solitons and Surface Anderson Localizations of Light}
\author{Viola Folli$^{1,3}$, Claudio Conti$^{2,1,3}$}
\affiliation{
$^1$Department of Physics, University Sapienza, Piazzale Aldo Moro, 5, 00185, Rome (IT)\\
$^2$Department of Molecular Medicine, University Sapienza, Viale Regina Elena, 324, 00185, Rome (IT)\\
$^3$ISC-CNR, Dep. Physics, Univ. Sapienza, Piazzale Aldo Moro 5, 00185, Rome (IT)}
\email{claudio.conti@roma1.infn.it}
\date{\today}

\begin{abstract}
Self-induced transparency pulses propagating in a random medium 
embedded in a two-level system can 
transfer energy to localized Anderson states. This allows the onset of two-level laser-like action.
\end{abstract}


\maketitle
\begin{figure}[t]
\includegraphics[width=\columnwidth]{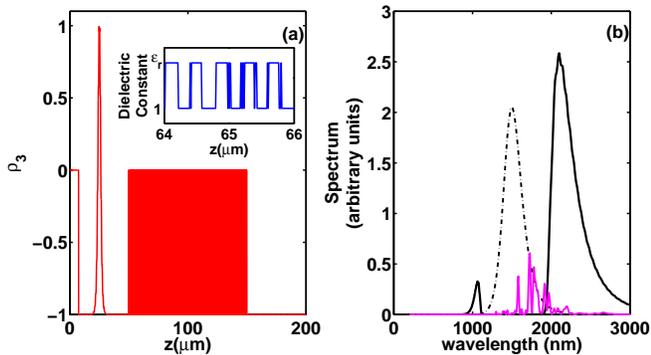}
\caption{(Color online) (a) The simulation region is 200$\mu$m long, the homogeneous two-level medium extends 
from $7.5\mu$m to $50\mu$m. The disordered two-level medium ($\epsilon_r=1$) extends
 from 50$\mu$m to 150$\mu$m; the initial population is equal to $\rho_{30}=-1$ and randomly alternated by a dielectric non-resonant medium with a different refractive index ($\rho_{30}=0$, $\epsilon_r=11$), 
as shown in the inset. A snapshot of the SIT pulse is shown before impinging on the structure; (b) 
SIT pulse spectrum for $\tau=100$fs (dashed line); spectrum of a transmitted short pulse
{\it without} the resonant medium showing the photonic-band gap (continuous line, $\gamma=0$);  transmitted spectrum {\it in absence} of 
resonant medium with $\gamma=0.5$ showing the disorder induced resonances in the forbidden band.}
\label{fig1} \end{figure}

The fact that population inversion cannot be achieved at thermal equilibrium prevents the 
possibility of observing laser action in a two level-system. \cite{EberlyBook}
This suggests to resort to an out-of-equilibrium regime
for a two-level laser, where light emission is amplified
by an ensemble of atoms pumped at the resonant frequency; at variance with
standard lasers, which adopt additional non-radiative decay levels that
unavoidably and drastically reduce the overall energetic efficiency 
(related is the case of Mollow gain \cite{Kaiser08}). 
A route is open by the well known Self-Induced Transparency process (SIT)
\cite{Kozlov97, dodd, mccall2,Malomed01,mantsyzov04,melnikov05,Rosanov10}: a short optical pulse propagating 
in an otherwise absorbing medium such that the resonant two-level atoms are excited in its leading edge, and 
coherently decay in the trailing edge. This mechanism is described by a soliton solution of the relevant 
Maxwell-Bloch equations \cite{lamb, Maimistov90}, which, however,
does not mediate a laser action, but allows
the transport of a population inversion region in strongly absorbing media.
Finding a way to localize such a region may lead to a two-level laser-like action. 
\\Here we propose to use a random system that supports many overlapping electromagnetic resonances \cite{Leuzzi09, Conti08,Cao03,Wiersma08,Lag2010,Andreasen10,Ciuti10}, such that the probability of trapping the traveling population inversion is largely enhanced.
The following issues are to be considered: 
(1) the effect of randomness on the SIT soliton; 
(2) cavities are created by defects in an ordered structure with a band-gap; 
(3) the disorder-induced resonances must overlap with the SIT soliton spectrum.
\noindent Points (2) and (3) lead to use a one-dimensional photonic crystal (PhC):  
a little disorder is indeed sufficient to create Anderson states in a known spectral region \cite{John87,Conti08PhC}.
\\For the point (1) above, we first consider the interaction of a SIT pulse with low-contrast refractive-index modulations. 
We theoretically show that disorder doesn't destroy the pulse, 
which fluctuates around its mean position with amplitude linearly growing with the propagation distance. 
In a later part of the manuscript, through first-principle numerical simulations, we analyze a disordered PhC in the high-index contrast case; 
we numerically find that at high scattering strength, 
the SIT pulse is trapped by coupling with band-gap Anderson states in proximity of the random structure surface \cite{Szameit10}.
After the arrival of the SIT pulse, the random cavity starts emitting laser-like radiation with population inversion.
\\We stress that non-solitonic pulses do not propagate in the 
absorbing medium containing the disordered PhC: 
using SIT is an unavoidable strategy to excite the disordered system.
In this respect, the interaction of SIT solitons with Anderson localizations provides a valuable route for a two-level laser-like action, while also being a fascinating regime for investigating the coupling
between two forms of light localization respectively due to nonlinearity and disorder \cite{Swartz07, Kivshar10, Conti10, ContiLeuzzi10}. 
\\\textit{Random SIT ---}  We first  address the role of a random perturbation  \cite{Folli10,Kozlov00} on a two-level $2\pi$ pulse by the reduced Maxwell-Bloch (MB) equations. 
In a later section, we discuss the non-perturbative regime by resorting to 
first-principle numerical simulations by a parallel Finite Difference Time Domain (FDTD) code \cite{ziolkowski}.
\noindent The MB equations with retarded time $\xi=t-z/v_g$ and space $z$ ($v_g$ is the velocity of the SIT pulse) read as 
\begin{eqnarray}
\label{siteqA}
\left[\frac{\partial }{\partial z}+\left( \frac{1}{c}-\frac{1}{v_g}\right)\frac{\partial}{\partial \xi}-i\mu \right]A=kB+S_1 \\
\label{siteqB}
\partial_{\xi} B=\frac{AN}{2}+S_2\text{,}\,\partial_{\xi} N=-\left(B^*A+BA^* \right)+S_3,
\end{eqnarray}
being:  $A$ the (slowly-varying) electric field, $B$ the atomic polarization and $N$ the population invertion,  $S_{1,2,3}$ the perturbations, $\mu$ the detuning from resonant frequency $\omega_0$,  $k=(4\pi\omega_0 N d^2)/c\hbar$, 
$d$ is the dipole moment, and $c$ the vacuum light velocity. 
 Let $\vec{A}=\vec{A}_s+\vec{A}_1$ with $\vec{A}_1$ a perturbation and 
$\vec{A}_s=\left(A_s, B_s, N_s\right)$ the soliton: 
\begin{equation}
\label{soliton}
\begin{array}{l}
\label{As}
\displaystyle A_s=u\mathcal{E}=\frac{2\beta\mathcal{E}}{\cosh\left( w\right)}\\
\label{Bs}
\displaystyle B_s=p\mathcal{E}=\frac{\beta^2}{\beta^2+\Omega^2}\left[\tanh \left( w\right)-\frac{i\Omega}{\beta}\right]\frac{\mathcal{E}}{\cosh\left( w\right)}\\
\label{Ns}
\displaystyle N_s=n=-1+\frac{2\beta^2}{\beta^2+\Omega^2}\frac{1}{\cosh^2\left( w\right)}
\end{array}
\end{equation}
where $w=\beta\left( \xi-X\right)$ and $\mathcal{E}=\exp\{-i\Omega(\xi-X)+i\theta\}$, $\mu\equiv\Omega k/(\beta^2+\Omega^2)$. 
$\beta$, $X$, $\Omega$, $\theta$ are the amplitude, position, detuning, and the phase of the soliton, being  
$1/v_g=1/c+k/(2\beta^2+2\Omega^2)$.
The linearized system is
\begin{equation}
\label{linearized}
\left(\partial_z \vec{A}_1\right)\cdot\mathcal{I}_1=\mathcal{L}_1(\vec{A}_1)+\vec{S}_p,
\end{equation}
where $\mathcal{I}_1=diag(1,0,0)$, $\vec{S}_p$ is the perturbation and
\begin{equation}
\mathcal{L}_1\left(\vec{A}_1\right)=\left(
\begin{array}{l}
 i\mu A_1-\left(\frac{1}{c}-\frac{1}{v_g}\right)\partial_\xi A_1+kB_1\\
 -\partial_\xi B_1+\frac{1}{2}\left(A_1N_s+A_sN_1\right)\\
 \partial_\xi N_1+2\Re(B^*_{s}A_1+B^*_1 A_s)
\end{array}\right)\text{.}
\end{equation}
Taking the SIT pulse resonant with the medium ($\Omega=\mu=0$), 
we derive equations for its parameters by first introducing the auxiliary functions \cite{Folli10}
$\vec{f}_X=\partial_{X}\vec{A}_s$, $\vec{f}_\theta=\partial_{\theta}\vec{A}_s$, $\vec{f}_\beta=\partial_{\beta}\vec{A}_s$ and $\vec{f}_\Omega=\partial_{\Omega}\vec{A}_s$, and letting
\begin{equation}
\label{A_1}
\vec{A}_1=\vec{f}_X \delta X+\vec{f}_\beta\delta\beta+\vec{f}_{\theta}\delta\theta+\vec{f}_\Omega\delta\Omega+\vec{a}_R\text{,}
\end{equation}
where $\delta X(z)$, $\delta\theta(z)$, $\delta\beta(z)$ and $\delta\Omega(z)$ are the time-dependent perturbations to soliton parameters and $\vec{a}_R$ is the radiation term ($\vec{a}_R=0$ hereafter, as it rapidly spreads and is absorbed.
\noindent We let $\vec{S}_p=\left(\left[V_B(z)p+iV_A(z)u\right]\mathcal{E}, 0, 0\right)$, such that 
$\langle V_{A,B}(z)V_{A,B}(z')\rangle=\langle V^{2}_{A,B}\rangle\delta(z-z')$,
where $V_{A,B}$ are the electric field and polarization perturbation.
Letting the adjoint functions\cite{Folli10} $\vec{\hat{f}}_{\theta}=i \vec{f}_\beta$,$\vec{\hat{f}}_\beta=-i \vec{f}_\theta$, $\vec{\hat{f}}_{\Omega}=-i \vec{f}_X$, $\vec{\hat{f}}_{X}=i \vec{f}_{\Omega_X}$, such that $(\vec{\hat{f}}_a,\vec{f}_b)=\mathcal{N}_a\delta_{a,b}$, with $a$ and $b$ two parameters in ($X$, $\Omega$, $\theta$, or $\beta$); we have $\mathcal{N}_\theta=\mathcal{N}_\beta=-1/(3\beta^2)$ and $\mathcal{N}_X=\mathcal{N}_\Omega=-2\pi\beta-1/(3\beta)$. 
Projecting over $\vec{\hat{f}}_k$ we get 
\begin{equation}
\label{ODE2}
\delta\dot{X}=-k/\beta^3\delta\beta+V_B(z)/2\beta^2\text{,}
\delta\dot{\theta}=-k/\beta^2\delta\Omega+V_A(z)
\end{equation}
and $\delta\dot{\beta}=\delta\dot{\Omega}=0$, which gives
\begin{equation}
\label{deltax}
\langle\delta X(z)^2\rangle=\displaystyle\frac{\langle V_B^2\rangle z}{4\beta^4}.
\label{fluct}
\end{equation}
Eq.~(\ref{fluct}) describes the random fluctuations of SIT solitons in the limit of small uncorrelated disorder, and states that they grow linearly with the propagation distance, and decay when increasing of the soliton power: as the velocity of the soliton is reduced, its random walking becomes more pronounced.
This result describes an index perturbation with weak index-contrast,
as shown below in figure \ref{fig2}a, figure  \ref{fig3}a,b and figure \ref{fig3}c for $\epsilon_{r1}$ and $\epsilon_{r2}$,
in the comparison with FDTD simulations.
Specifically, eq.(\ref{fluct}) predicts that slower SIT solitons are those
mostly affected by disorder, and hence they are expected to interact more effectively
with Anderson states also in the high index contrast regime.
As we numerically show in the following (figures  \ref{fig2}b, \ref{fig3}c for $\epsilon_{r3}$, \ref{fig4} and \ref{fig5}), as the strength of disorder increases 
(when the theoretical analysis reported above is not expected to the valid), this leads to localization and laser-like action.
\\\textit{Numerical Results ---} Following Ref.~\cite{ziolkowski}, we numerically solve the one-dimensional full MB equations by a parallel FDTD algorithm and study the evolution of the SIT pulse in a non-linear disordered SIT medium; we denote as $\rho_3$ the population inversion of the FDTD equations 
(corresponding to $N$ in Eq.(\ref{siteqB}) when adopting the rotating wave approximation) and $E$ is the real-valued electric field, whose slowly-varying complex envelope is $A$ in (\ref{siteqA}). Note that we use a different notation for the field here because the FDTD equations are more general that the reduced MB equations 
(\ref{siteqA}, \ref{siteqB}) \cite{ziolkowski}. 
\\As sketched in figure \ref{fig1}, the SIT pulse propagates into the grid from the left boundary $(z=0)$,
from vacuum ($z<7.5\mu$m) to a homogeneous SIT medium (atom density $N_{atoms}=10^{24}$m$^{-3}$);  its carrier frequency is resonant 
with the medium $\omega_0=2\pi f_0$, with $f_0=2\times 10^{14}$s$^{-1}$, 
and the pulse duration $\tau$ is chosen to satisfy the
$2\pi$ area theorem.
The relaxation time of the density matrix equations are $T_1=T_2=1.0\times 10^{-10}$s ($T_{1,2}>>\tau$). 
The initial population inversion in the SIT medium is $\rho_{30}=-1$, while in the dielectric layers $\rho_{30}=0$
and the two-level atoms are not present. We neglect absorption in the dielectric layers because much smaller than in the resonant two-level system.
The input pulse, with peak $E_0=(2\hbar/d)\beta$, 
enters the two-level medium, initially set in the ground state  $\rho_{30}=-1$, and
generates the SIT soliton, which, after propagating in the homogeneous resonant region, interacts with the random structure as shown in Fig.~\ref{fig1}.
The total length of the structure is 200$\mu$m, the random active medium extends from 50$\mu$m to 150$\mu$m,
where we add a fixed degree of disorder by inserting slices of a non absorbing medium with relative permittivity $\epsilon_r$.
The disordered structure is created by introducing random layers in the homogeneous two level medium. 
The degree of disorder is quantified by a parameter $\gamma$: 
the relative permittivity distribution has a square-wave profile from $1$ to $\epsilon_r$ (inset in figure 1a) 
given by the sign of the function $\sin(2\pi z/d+2\pi \gamma \zeta)$,
with $\zeta$ a uniform deviate in $\left[0,1\right]$ extracted in each point of the grid.
If $\gamma=0$ an ordered periodical structure is attained: 
the dielectric layers are equidistant, with constant width $200$nm and period $400$nm ($d=400$nm) 
and displays a band-gap centered at $\omega_0$ overlapped with the spectrum of the SIT soliton (see figure \ref{fig1}b). 
For $\gamma=0$, increasing $\epsilon_r$ creates a one dimensional band-gap structure;
for $\gamma>0$, increasing $\epsilon_r$ enforces the effect of disorder, and as shown in figure \ref{fig2}b in the absence of the two-level system; the 
localized states first appears in proximity of the band-edge of the ordered structure.
We use this approach to selectivity create a distribution of Anderson states with a spectrum superimposed to that of the input SIT pulse.
In the following, the described disordered structure is embedded in a two-level system and SIT solitons are launched in the
homogeneous region ($z<50\mu$m) and then interact with the random system ($z>50\mu$m). Non-solitonic pulses are rapidly absorbed and do not propagate.\\
\begin{figure}[t]
\includegraphics[width=\columnwidth]{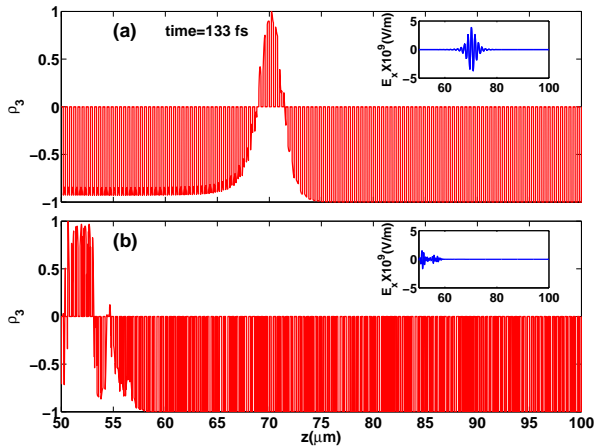}
\caption{
(Color online) SIT population inversion $\rho_3$ in (a) the low-index case ($\epsilon_r=1.5$, $\gamma=0$,
analogous results for $\gamma>0$) and in (b) the high index case ($\epsilon_r=11$, $\gamma=1$), where surface 
Anderson states are excited. The insets show the corresponding $E$ profiles.
$\rho_3$ appears periodically modulated as $\rho_3=0$ in the dielectric layers with $\epsilon_{r}>1$.}
\label{fig2} \end{figure}
\begin{figure}[t]
\includegraphics[width=\columnwidth]{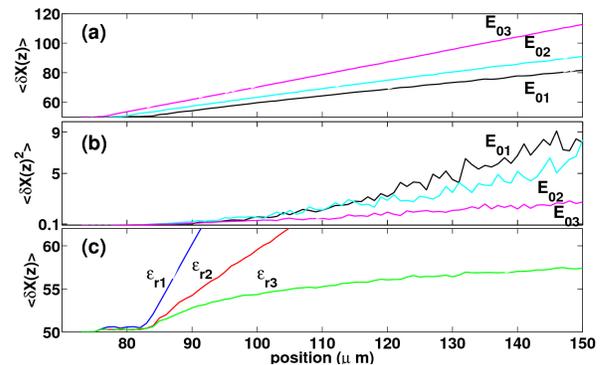}
\caption{(Color online) (a) SIT pulse trajectories calculated for $\gamma=0.1$, $\epsilon_r=6$,
for three different initial pulse 
amplitudes $E_{01}=4.2186\times10^9V/m$, $E_{02}=4E_{01}$, $E_{03}=8E_{01}$; (b) corresponding standard deviations; (c) SIT pulse trajectories for three relative permittivities $\epsilon_{r1}=2.5$, $\epsilon_{r2}=6$, $\epsilon_{r3}=11$ when $E_{0}=E_{01}$.
Note that the curves for $\epsilon_{r1,2}$ are cut 
(they proceed as straight lines) for comparison with the case $\epsilon_{r3}.$
The results are averaged over $100$ realizations of the disorder.
}
\label{fig3} \end{figure}
Figure~\ref{fig2} shows a temporal snapshot of the population inversion distribution in low index ($\epsilon_r=1.5$, $\gamma=0$) 
ordered case and in the high index disordered case ($\epsilon_r=11$, $\gamma=1$):
in the former case, the wave propagates with limited distortion and the inversion population is transported through the medium,
in agreement with what expected by the perturbation theory reported above
(analogous results for low index-contrast $\epsilon_r\cong1$ are obtained for $\gamma>0$, see figure \ref{fig3}b,c); in the latter case, the spatial distribution of the inversion population localizes in proximity of the input face of the sample. 
This shows that surface Anderson states are involved in the process \cite{Szameit10}; as we will report in future work
multiple SIT pulses can be employed to enlarge the trapped inversion region.

Panels a,b in Fig.~\ref{fig3} show the trajectories and the corresponding standard deviation of the SIT pulse, 
calculated for different input peak values and for a fixed strength of disorder. 
In the low index contrast disordered case, following the previous theoretical analysis (see Eq.~(\ref{deltax})), as the initial peak value increases, the soliton propagates
faster in the structure, and the slope in Fig.~\ref{fig3}a increases and the corresponding fluctuations are reduced (Fig.~\ref{fig3}b).
Conversely, slow solitons perform a more pronounced random walk in the disordered structure.
As the index contrast is increased (beyond the regime of validity of the perturbational approach reported above), 
this allows a localization processes as shown
for $\epsilon_{r3}$ in  Figure~\ref{fig3}c displaying soliton trajectories for different strengths of the disorder (fixed $\gamma$ and increasing $\epsilon_r$).
In the cases $\epsilon_{r1}$ and $\epsilon_{r2}$ the trajectories proceeds as straight lines, denoting a propagating pulse, conversely for $\epsilon_{r3}$ the
trajectory bends and the soliton slows down and is trapped.
In the other words, in the low-index contrast case,
the soliton displays a weakly perturbed motion; as the strength of disorder increases
(when increasing $\epsilon_r$ with fixed $\gamma$), its trajectory becomes more fluctuating, 
until the wave gets localized; correspondingly the system starts emitting laser-like radiation,
as a population inversion region is formed in correspondence of an optical cavity.

Figure \ref{fig4} shows the output of the device for low and high index contrast for a fixed $\gamma$.
For a small $\epsilon_r$ the input SIT pulse in \ref{fig4}a  is transmitted as shown in \ref{fig4}b,
where the pulse exits at about $0.5$ps.
On the contrary, for high $\epsilon_r$, 
the pulse is trapped in the structure and, 
after a transient, the output (in Fig.\ref{fig4}c) 
corresponds to a laser-like emission from the the two-level system.
Note the difference in the vertical scale: for Fig.\ref{fig4}b all the energy is transmitted, 
while for Fig.\ref{fig4}c, it is partially reflected (see also figure \ref{fig5}) and partially trapped
in the light-emitting localized states.
In figure \ref{fig5}a, we show the reflected temporal signal, comprising of the portion of the SIT pulse, which is not trapped in the disordered structure and is reflected (large peak around $0.3$ps), and the subsequent emission from the population inversion trapped in the disorder medium; 
its spectrum in \ref{fig5}b displays characteristic peaks signaling 
the excited Anderson localizations with frequencies in the forbidden gap. The energy trapped in the localized modes decay
with their characteristic lifetimes (of the order of $0.5$ps).
\begin{figure}
\includegraphics[width=\columnwidth]{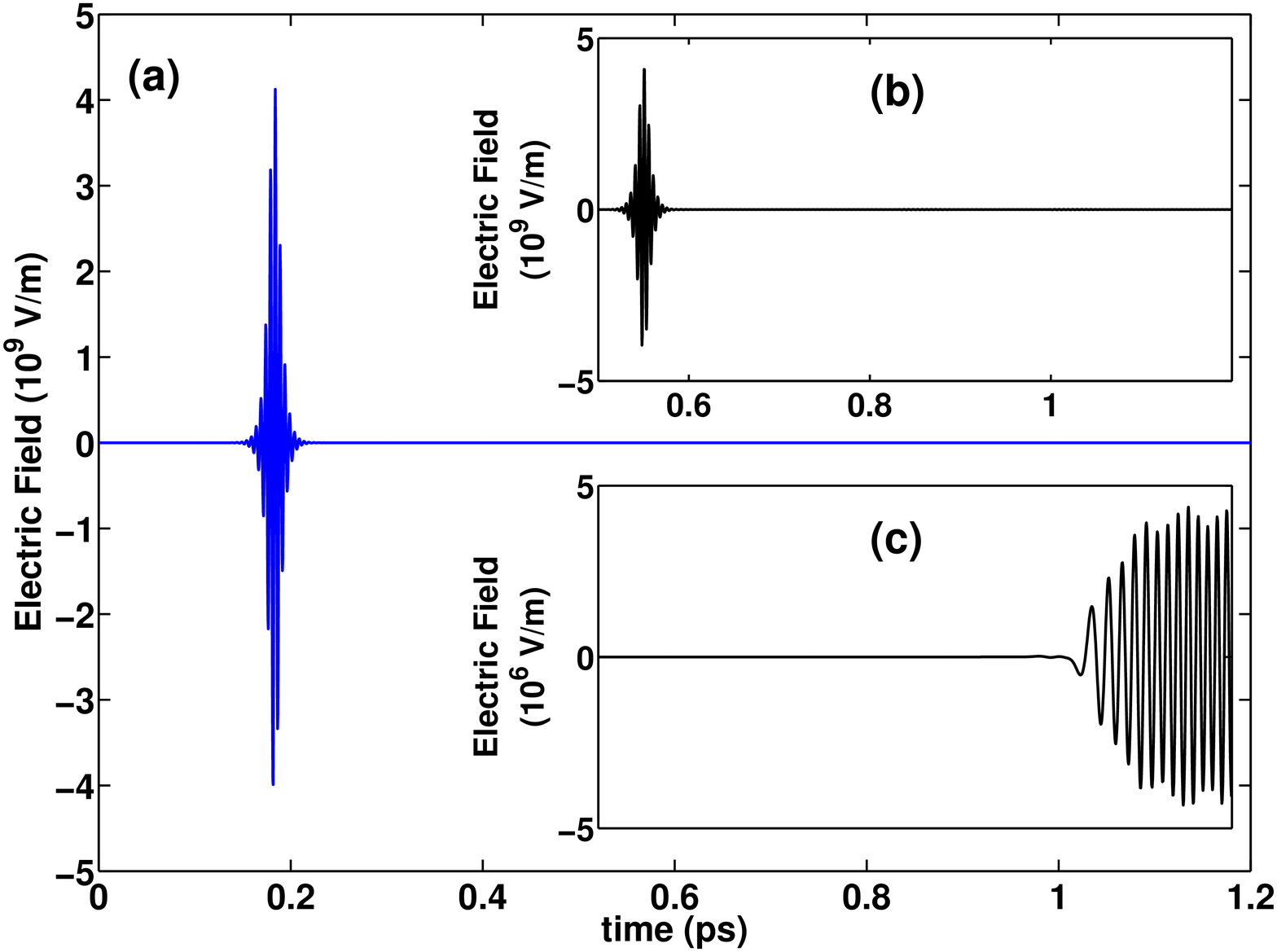}
\caption{(Color online) (a) Temporal profile of the input SIT-pulse with $\tau=100$fs and $E_0=4.2186\times 10^9$V$/$m;
(b) transmitted pulse without disordered structure; (c) emitted signal 
for an highly scattering medium ($\epsilon_r=11$, $\gamma=1$)
The horizontal scale in panels b,c is overlapped with panel a. }
\label{fig4} \end{figure}
\begin{figure}
\includegraphics[width=8.3cm]{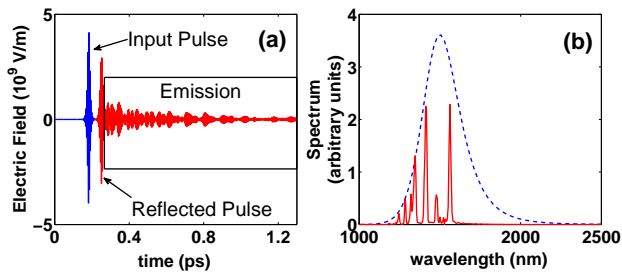}
\caption{(Color online) (a) Temporal profiles of the reflected electric field in the highly scattering case in Fig\ref{fig4}c,
including the input pulse, its reflected fraction, and the subsequent two-level laser emission; 
(b) corresponding spectrum (continuous line) compared with the input SIT pulse (dashed line). }
\label{fig5}
\end{figure}

\noindent {\it Conclusions ---} 
We investigated the interaction of a SIT soliton with Anderson localizations through theory and parallel Maxwell-Bloch simulations.
We have shown that an increasing scattering strength
of disorder progressively slows down the soliton, up to blocking the pulse in the random system.
This process is accompanied by the excitation of modes in a disordered cavity in the presence of population inversion. This results into a two-level laser-like action.
The interplay between various forms of light localizations, namely solitons and Anderson states,
can hence lead to novel processes of light-matter interactions in complex systems.

We acknowledge support from the CINECA-ISCRA parallel computing initiative.
The research leading to these results has received funding from the
European Research Council under the European Community's Seventh Framework Program 
(FP7/2007-2013)/ERC grant agreement n.201766.


\end{document}